\documentclass[a4paper]{article}
\usepackage[T1]{fontenc}
\usepackage{float}
\usepackage{multirow}
\usepackage{makecell}
\usepackage{amsthm}
\usepackage{amsfonts}
\usepackage{amssymb}

\usepackage{graphicx}
\usepackage{multirow}
\usepackage[hidelinks]{hyperref}
\usepackage{authblk}
\usepackage{booktabs,caption}
\usepackage[flushleft]{threeparttable}
\usepackage{lscape}
\usepackage{rotating}
\usepackage{pgfplots}

\mathchardef\mhyphen="2D

\usepackage[backend=biber, style=apa]{biblatex}
\providecommand{\keywords}[1]
{
  \small	
  \textbf{\textit{Keywords---}} #1
}

\title{OpenCitations Meta}

\author[1,3]{Arcangelo Massari\thanks{arcangelo.massari@unibo.it}\href{https://orcid.org/0000-0002-8420-0696}{\includegraphics{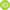}}}
\author[2]{Fabio Mariani\thanks{fabio.mariani@leuphana.de}\href{https://orcid.org/0000-0002-7382-0187}{\includegraphics{img/orcid.png}}}
\author[1,3]{Ivan Heibi\thanks{ivan.heibi2@unibo.it}\href{https://orcid.org/0000-0001-5366-5194}{\includegraphics{img/orcid.png}}}
\author[1,3]{Silvio Peroni\thanks{silvio.peroni@unibo.it}\href{https://orcid.org/0000-0003-0530-4305}{\includegraphics{img/orcid.png}}}
\author[1,4]{David Shotton\thanks{david.shotton@opencitations.net}\href{https://orcid.org/0000-0001-5506-523X}{\includegraphics{img/orcid.png}}}
\affil[1]{Research Centre for Open Scholarly Metadata, Department of Classical Philology and Italian Studies, University of Bologna, Bologna, Italy}
\affil[2]{Institute of Philosophy and Sciences of Art, Leuphana University, Lüneburg, Germany}
\affil[3]{Digital Humanities Advanced Research Centre (/DH.arc), Department of Classical Philology and Italian Studies, University of Bologna, Bologna, Italy}
\affil[4]{Oxford e-Research Centre, University of Oxford, Oxford, United Kingdom}

\date{}

\begin{document}

\maketitle

\begin{abstract}
OpenCitations Meta is a new database for open bibliographic metadata of scholarly publications involved in the citations indexed by the OpenCitations infrastructure, adhering to Open Science principles and published under a CC0 license to promote maximum reuse. It presently incorporates bibliographic metadata for publications recorded in Crossref, DataCite and PubMed, making it the largest bibliographic metadata source using Semantic Web technologies. It assigns new globally persistent identifiers (PIDs), known as OpenCitations Meta Identifiers (OMIDs) to all bibliographic resources, enabling it both to disambiguate publications described using different external PIDS (e.g., a DOI in Crossref and a PMID in PubMed), and to handle citations involving publications lacking external PIDs. By hosting bibliographic metadata internally, OpenCitations Meta eliminates its former reliance on API calls to external resources and thus enhances performance in response to user queries. Its automated data curation, following the OpenCitations Data Model, includes deduplication, error correction, metadata enrichment and full provenance tracking, ensuring transparency and traceability of data and bolstering confidence in data integrity, a feature unparalleled in other bibliographic databases. Its commitment to Semantic Web standards ensures superior interoperability compared to other machine-readable formats, with availability via a SPARQL endpoint, REST APIs and data dumps.
\end{abstract}


\section{Introduction}\label{introduction}
OpenCitations is an independent not-for-profit infrastructure organization for open scholarship dedicated to publishing open bibliographic and citation data using Semantic Web technologies. OpenCitations stores and manages information about scholarly citations, i.e., the conceptual links connecting a citing entity with a cited entity, in the OpenCitations Indexes. Hitherto, there have been four OpenCitations Indexes: COCI (\url{https://opencitations.net/index/coci}), the OpenCitations Index of Crossref open DOI-to-DOI Citations  \parencite{heibi_software_2019}; POCI (\url{https://opencitations.net/index/poci}), the OpenCitations Index of PubMed open PMID-to-PMID citations; DOCI (\url{https://opencitations.net/index/doci}), the OpenCitations Index of DataCite open DOI-to-DOI citations; and CROCI (\url{https://opencitations.net/index/croci}), the Crowdsourced Open Citations Index \parencite{heibi_crowdsourcing_2019}.

Building upon the foundational work of OpenCitations, which has been instrumental in publishing open citation data, OpenCitations Meta emerges as a complementary database. Although the OpenCitations Indexes focus primarily on scholarly citations, OpenCitations Meta manages the metadata of the citing and cited academic publications in-house. With this backdrop, it becomes imperative to understand the benefits OpenCitations Meta brings to its users, especially when compared to other infrastructures.

First, OpenCitations Meta employs a robust provenance model \parencite{daquino_opencitations_2020}, grounded in the PROV Ontology \parencite{lebo_prov-o_2013}, a standard endorsed by the World Wide Web Consortium (W3C). This model fosters transparency and traceability, ensuring that every bibliographic resource within OpenCitations Meta is enriched with detailed provenance information. This encompasses specifics about the agents involved in the data's creation, modification, deletion, and merging, the temporal markers of these operations, and the primary sources from which the information was derived. By housing this provenance data in the Resource Description Framework (RDF), it aligns seamlessly with other semantic web tools and technologies. Such a meticulous approach to provenance fortifies trust in the data's authenticity, granting users the capability to trace both the origins and subsequent modifications of any information. As we explore in Section \ref{sec:related_works}, this depth of transparency sets a benchmark that many other infrastructures have yet to achieve.

Second, OpenCitations Meta incorporates a sophisticated change-tracking mechanism. This feature is crucial for maintaining the accuracy and reliability of data, especially in dynamic environments where information is continually updated. Every modification, addition, deletion or merge within the data set is recorded. The change-tracking system is built on the principles of the PROV Ontology, ensuring that every change is associated with its respective provenance. Users can access the entire history of a data item, making it easy to understand its evolution over time. Notably, no other bibliographic index offers such a granular view of resource evolution in RDF format.

The third standout feature of OpenCitations Meta is its commitment to semantic data, which plays a pivotal role in complying with the FAIR principles \parencite{wilkinson_fair_2016}. Compared to other machine-readable formats such as CSV, RDF offers a superior level of interoperability. Although CSV files are indeed machine readable, they lack the semantic richness and interconnectedness of RDF. CSV files are essentially flat tables, and although they can represent data, they do not capture the relationships between data entities in the way RDF does. RDF's graph-based nature allows for a more intricate web of data, where entities and their relationships are clearly defined and linked. This ensures a higher level of interoperability, as systems can understand not just the data, but also the context and relationships surrounding that data.

Moreover, although the coverage of the OpenCitations Indexes has approached parity with that of commercial proprietary citation indexes \parencite{martin-martin_coverage_2021}, there have previously been three outstanding issues not formerly addressed by OpenCitations.

First is the previously poor temporal performance of the OpenCitations' services, in particular API operations returning basic bibliographic metadata of citing and cited resources. This is because the OpenCitations Indexes themselves have hitherto contained only citation-related metadata (citations being treated as First Class data entities with their own metadata), but have not held bibliographic metadata relating to the citing and cited entities (title, authors, page numbers, etc.). Rather, those metadata have hitherto been retrieved on-the-fly by means of explicit API requests to external services such as Crossref, ORCID and DataCite.

Second is citation disambiguation. Sometimes, bibliographic resources will have been assigned multiple identifiers, such as a DOI \textit{and} a PMID, by separate external services (e.g., Crossref and PubMed). In such cases, the same citation may be multiply represented in different ways depending on the data source. For example, OpenCitations will describe in COCI a citation between two publications using metadata derived from Crossref as a DOI-to-DOI citation, and in POCI the same citation using metadata derived from PubMed as a PMID-to-PMID citation. This duplication poses problems when counting the number of ingoing and outgoing citations of each document, a crucial statistic for libraries, journals, and scientometrics studies. Use of OpenCitations Meta permits us to deduplicate such citations and solve the problems that such duplication would otherwise cause. 

Third, the assignment of globally persistent identifiers to documents is not universal practice across all scholarly fields. Gorraiz et al. \parencite*{gorraiz_availability_2016} demonstrated that the Natural and Social Sciences communities adopt DOIs to a much greater extent than the Arts and Humanities community. From that research, carried out on Scopus and the Web of Science Core Collection, it emerged that almost 90\% of the publications in the Sciences and Social Sciences are associated with a DOI, whereas in the Arts and Humanities, that figure is only 50\%. In addition, concerning the Humanities, citations of ancient primary sources lacking DOIs (e.g., Aristotle) are required in many fields (e.g., in History). If a document has no identifier, its metadata do not meet the ``identifiability'' standard set forth by the FAIR principles \parencite{wilkinson_fair_2016}, which stipulate that scholarly digital research objects should be findable, accessible, interoperable, and reusable. A globally unique and persistent identifier is critical to make metadata findable and accessible. Moreover, a bibliographic resource without an identifier prevents citations involving it from being described adhering to the FAIR principles. This is the reason why, according to the Open Citation Definition \parencite{peroni_open_2018} governing the population of OpenCitations Indexes, any two entities linked by an indexed citation must both be identified by a persistent identifier coming from the same identifier scheme, for example both with DOIs, or both with PMIDs. For example, COCI \parencite{heibi_software_2019} only stores citation information where the citing and cited entities are described in Crossref and both have DOIs. Citations involving publications lacking DOIs or other recognized PIDs have hitherto been excluded from the OpenCitations citation indexes. 

\begin{figure}[H]
    \centering
    \includegraphics[width=\textwidth]{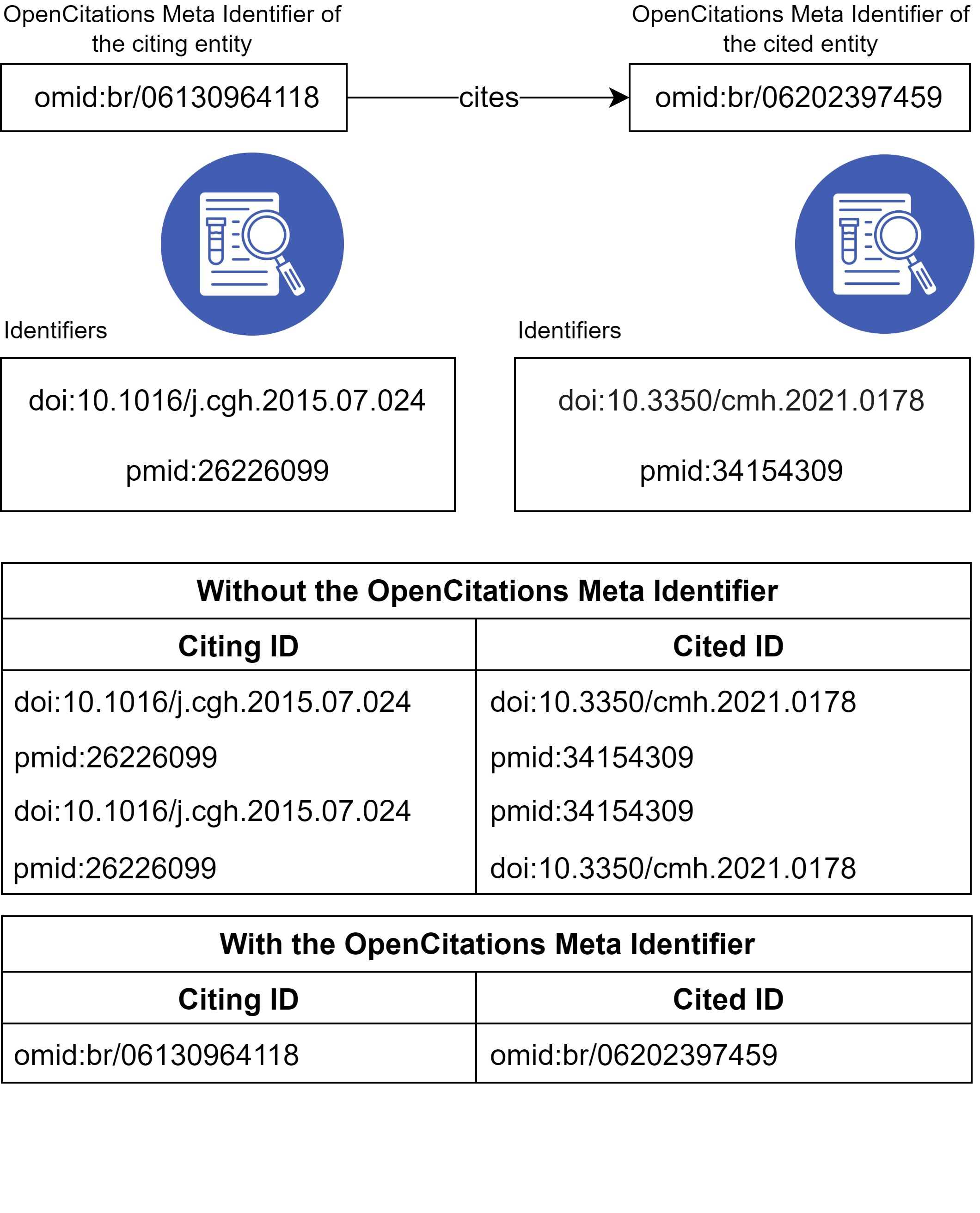}
    \caption{If a document is described by multiple identifiers, e.g., a DOI from Crossref and a PMID from Pubmed, the citations involving it may be described in multiple ways, creating an ambiguity and deduplication problem. Use of the OpenCitations Meta Identifier solves this issue by acting as a proxy between different external identifiers.}
    \label{fig:citations_deduplication}
\end{figure}

But now, OpenCitations Meta solves the problems posed by bibliographic resources identified by multiple identifiers and also bibliographic resources that lack persistent identifiers, by associating a new globally persistent identifier to each document described in OpenCitations Meta - an OpenCitations Meta Identifier (OMID). In this way, all citations can be represented as OMID-to-OMID citations (Fig. \ref{fig:citations_deduplication}). By providing a unique identifier for every entity stored in OpenCitations Meta, the entity's OMID acts as a proxy between different external identifiers used for each entity, enabling disambiguation. Moreover, OpenCitations Meta can contain metadata for all scholarly publication, each identified by an OMID, without the mandatory need for an external persistent identifier to be provided by the source of the metadata.

Thus, thanks to OpenCitations Meta, metadata for \textit{all} scholarly publications can now be stored by OpenCitations, and citations linking \textit{all} such publications can be included within a new inclusive OpenCitations Index, of which the other indexes (COCI, DOCI, POCI, etc.) will be subindexes, according to the various input sources of the citation information.

Over the past 3 years, to address the issues mentioned above, we have developed and tested the software we are now using to create this new bibliographic metadata collection, namely OpenCitations Meta, which we launched in December 2022. The software supporting this database is open source, and available at \url{https://github.com/opencitations/oc_meta}. The metadata exposed by OpenCitations Meta includes the basic bibliographic metadata describing a scholarly bibliographic resource. In particular, it stores all known bibliographic resource identifiers for the bibliographic resource (e.g., DOIs, PMIDs, ISSNs, and ISBNs), the title, type, publication date, pages, the venue of the resource, and the volume and issue numbers where the venue is a journal. In addition, OpenCitations Meta contains metadata regarding the main actors involved in the publication of each bibliographic resource, i.e., the names of the authors, editors, and publishers, each including their own persistent identifiers (e.g., ORCIDs) where available. It is our intention to add additional metadata fields (e.g., authors' institutions and funding information) at a later date.

The process of generating OpenCitations Meta can be divided into two steps. The first step involves the curation of the input data. The curatorial procedure concerns the automatic correction of errors, the standardization of the data format, and the deduplication of separate metadata entries for the same item. The deduplication process is based only on identifiers. This approach favors precision over recall: for instance, people are deduplicated only if they have an assigned ORCID, and never by other heuristics. After the normalization and deduplication stages, each entity is assigned an OpenCitations Meta Identifier (OMID), whether or not it already has an external persistent identifier (e.g., DOI, PMID, ISBN). 

The second step in populating OpenCitations Meta involves converting the raw input data into RDF (Linked Open Data format) compliant with the OpenCitations Data Model (OCDM) \parencite{daquino_opencitations_2020}, to enable querying such data via SPARQL. During this process, great attention is given to provenance and change-tracking: every time an entity is created, modified, deleted or merged, such changes are recorded in RDF, and are characterized by their dates of creation, primary sources, and responsible agents.

As we delve deeper into explaining the intricacies of OpenCitations Meta, we will employ various Semantic Web terminologies. For readers unfamiliar with these concepts, especially in the context of semantic publishing, the blog post ``Linked Data 101'' by David Shotton provides an excellent primer \parencite{shotton_linked_2013}.

The rest of the paper is organized as follows. Section \ref{sec:related_works} reviews other data sets represented using Semantic Web technologies. Subsequently, in Section \ref{sec:methodology}, the methodological approach adopted to produce OpenCitations Meta is presented in detail, starting with the curatorial phase (\ref{sec:dedup}), then describing error correction (\ref{sec:error_proofing}), moving to an explanation of the data translation to RDF according to the OCDM (\ref{subsec:semantic_mapping}), and concluding with a description of the production of the RDF provenance and change-tracking data  (\ref{subsec:provenance}). Section \ref{sec:eval} provides some descriptive statistics regarding the current OpenCitations Meta data set. Finally, Section \ref{sec:discussion} discusses some present limitations of OpenCitations Meta, and a consideration of where OpenCitations Meta stands among similar scholarly data sets.

\section{Related works}\label{sec:related_works}

In this section, we will review the most important scholarly publishing data sets to which access does not require subscription, i.e., publicly available data sets holding scholarly bibliographic metadata. Since OpenCitations Meta uses Semantic Web technologies to represent data, special attention will be given to RDF data sets, namely Wikidata, Springer Nature SciGraph, BioTea, the OpenResearch Knowledge Graph and Scholarly Data. In addition, the OpenAIRE Research Graph, OpenAlex and Scholarly Data will be described, as they are the most extensive data sets in terms of the number of works, although they do not represent data semantically. 

OpenAlex \parencite{priem_openalex_2022} rose from the ashes of the Microsoft Academic Graph on January 1st 2022, and inherited all its metadata. It includes data from Crossref \parencite{hendricks_crossref_2020}, PubMed \parencite{maloney_pubmed_2013}, ORCID \parencite{haak_orcid_2012}, ROR \parencite{lammey_solutions_2020}, DOAJ \parencite{morrison_directory_2017}, Unpaywall \parencite{dhakal_unpaywall_2019}, arXiv \parencite{sigurdsson_future_2020}, Zenodo \parencite{european_organization_for_nuclear_research_zenodo_2013}, the ISSN International Centre\footnote{https://www.issn.org/}, and the Internet Archive's General Index\footnote{https://archive.org/details/GeneralIndex}. In addition, web crawls are used to add missing metadata. With over 240 million works\footnote{\url{https://docs.openalex.org/api-entities/works}}, OpenAlex is the most extensive bibliographic metadata data set to date. OpenAlex assigns persistent identifiers to each resource. In addition, authors are disambiguated through heuristics based on coauthors, citations, and other features of the bibliographic resources. The data are distributed under a CC0 license and can be accessed via API, web interface or downloading a full snapshot copy of the OpenAlex database.

The OpenAIRE project started in 2008 to support the adoption of the European Commission Open Access mandates \parencite{manghi_infrastructure_2010}, and it is now the flagship organization within the Horizon 2020 research and innovation program to realise the European Open Science Cloud \parencite{european_commission_directorate_general_for_research_and_innovation_realising_2016}. One of its primary outcomes is the OpenAIRE Research Graph, which includes metadata about scholarly outputs (e.g., literature, data sets and software), organizations, research funders, funding streams, projects, and communities, together with provenance information. Data are harvested from a variety of sources \parencite{grana_openaire_2017}: archives, e.g., ArXiv \parencite{sigurdsson_future_2020} Europe PMC \parencite{the_europe_pmc_consortium_europe_2015}, Software Heritage \parencite{abramatic_building_2018} and Zenodo \parencite{european_organization_for_nuclear_research_zenodo_2013}; aggregator services, e.g., DOAJ \parencite{morrison_directory_2017} and OpenCitations \parencite{peroni_opencitations_2020}; and other research graphs, e.g., Crossref \parencite{hendricks_crossref_2020} and DataCite \parencite{brase_datacite_2009}. As of June 2023, this OpenAIRE data set consisted of 232,174,001 research products\footnote{\url{https://explore.openaire.eu/search/find/research-outcomes}}. The deduplication process implemented by OpenAIRE takes into account not only PIDs but also other heuristics, such as the number of authors and the Levenstein distance of titles. However, the internal identifiers OpenAIRE associates with entities are not persistent and may change when the data are updated. Data of the OpenAIRE Research Graph can be accessed via an API and the Explore interface.  Dumps are also available under a Creative Commons Attribution 4.0 International Licence.

Semantic Scholar was introduced by the Allen Institute for Artificial Intelligence in 2015 \parencite{fricke_semantic_2018}. It is a search engine that uses artificial intelligence to select only papers most relevant to the user's search and to simplify exploration, e.g., by producing automatic summaries. Semantic Scholar sources its content via web indexing and partnerships with scientific journals, indexes, and content providers. Among those are the Association for Computational Linguistics, Cambridge University Press, IEEE, PubMed, Springer Nature, The MIT Press, Wiley, arXiv, HAL, and PubMed. As of June 2023, it indexes 212,605,886 scholarly works\footnote{\url{https://www.semanticscholar.org/}}. Authors are disambiguated via an artificial intelligence model \parencite{subramanian_s2and_2021}, associated with a Semantic Scholar ID, and a page is automatically generated for each author, which the real person can redeem. Semantic Scholar provides a web interface, APIs, and the complete data set is downloadable under the Open Data Commons Attribution Licence (ODC-By) v1.0.

Wikidata was introduced in 2012 by Wikimedia Deutschland as an open knowledge base to store in RDF data from other Wikimedia projects, such as Wikipedia, Wikivoyage, Wiktionary, and Wikisource \parencite{mora-cantallops_systematic_2019}. Due to its success, Google closed Freebase in 2014, which was intended to become ``Wikipedia for structured data'' and migrated it to Wikidata \parencite{tanon_freebase_2016}. Since 2016, the WikiCite project has contributed significantly to the evolution of Wikidata as a bibliographic database, such that, by June 2023, Wikidata contained descriptions of 39,864,447 academic articles\footnote{\url{https://scholia.toolforge.org/statistics}}. The internal Wikidata identifier referring to any entity (including bibliographic resources) is associated with numerous external identifiers, e.g., DOI, PMID, PMCID, arXiv, ORCID, Google Scholar, VIAF, Crossref funder ID, ZooBank and Twitter. The data are released under a CC0 license as RDF dumps in Turtle and NTriples. Users can browse them via SPARQL, a web interface and, as of 2017, via Scholia – a web service which performs real-time SPARQL queries to generate profiles on researchers, organizations, journals, publishers, academic works and research topics, while also generating valuable infographics \parencite{nielsen_scholia_2017}.

Although OpenAIRE Research Graph and Wikidata aggregate many heterogeneous sources, Springer Nature SciGraph \parencite{hammond_data_2017}, on the other hand, aggregates only data from Springer Nature and its partners. It contains entities concerning publications, affiliations, research projects, funders and conferences, totalling more than 14 million research products\footnote{\url{https://scigraph.springernature.com/explorer/data sets/data_at_a_glance/}}. There is no current plan to offer a public SPARQL endpoint, but there is the possibility to explore the data via a browser interface, and a dump is released monthly in JSON-LD format under a CC-BY license.

BioTea is also a domain-oriented data set, and represents the annotated full-text open-access subset of PubMed Central (PMC-OA) \parencite{garcia_biotea_2018} using RDF technologies. At the time of that 2018 paper, the data set contained 1.5 million bibliographic resources. Unlike other data sets, BioTea describes metadata and citations and defines the annotated full-texts semantically. Named-entity recognition analysis is adopted to identify expressions and terminology related to biomedical ontologies that are then recorded as annotations (e.g., about biomolecules, drugs, and diseases). BioTea data are released as dumps in RDF/XML and JSON-LD formats under the Creative Commons Attribution Non-Commercial 4.0 International license, and the SPARQL endpoint is currently offline.

A noteworthy approach is that adopted by the Open Research Knowledge Graph (ORKG) \parencite{auer_improving_2020}. Metadata are mainly collected either by trusted agents via crowdsourcing or automatically from Crossref. However, ORKG's primary purpose is not to organise metadata but to provide services. The main scope of these services is to perform a literature comparison analysis using word embeddings to enable a similarity analysis and foster the exploration and link of related works. To enable such sophisticated analyses, metadata from Crossref is insufficient; therefore, structured annotations on the topic, result, method, educational context and evaluator must be manually specified for each resource. The data set contains (as of June 2023) 25,680 papers\footnote{\url{https://orkg.org/papers}}, 5153 data sets, 1364  software and 71 reviews. Given the importance of human contribution to the creation of the ORKG data set, the platform keeps track of changes and provenance, although not in RDF format. The data can be explored through a web interface, SPARQL, and an API, and can also be downloaded under a CC BY-SA license.

ScholarlyData collects information only about conferences and workshops on the topic of the Semantic Web \parencite{nuzzolese_semantic_2016}. Data are modeled following the Conference Ontology, which describes typical entities in an academic conference, such as accepted papers, authors, their affiliations, and the organizing committee, but not bibliographic references. Up to June 2023, the data set stored information about 5678 conference papers. Such a data set is updated by employing the Conference Linked Open Data generator software, which outputs RDF starting from CSV files \parencite{gentile_clodg-conference_2015}. The deduplication of the agents is based only on their URIs using a supervised classification method \parencite{blomqvist_entity_2017}, and ORCIDs are added in a further step. This methodology does not address the existence of homonyms. However, this is a minor issue for ScholarlyData, since only a few thousand people are involved in the conferences being indexed. ScholarlyData can be explored via a SPARQL endpoint, and dumps are available in RDF/XML format under a Creative Commons Attribution 3.0 Unported license.

To conclude, we would like to point out that none of these other data sets mentioned above exposes change-tracking data and the related provenance information in RDF.

Table \ref{datasets_comparison} summarizes all the considerations made on each data set.

\begin{sidewaystable*}
\centering
\caption{Open scholarly data sets ordered by the number of contained research entities, and compared regarding change-tracking, provenance, disambiguation method, presence of an internal ID, accessibility, and data usage license.}
\label{datasets_comparison}
\resizebox{\textwidth}{!}{%
\begin{tabular}{llllllll}
\textbf{Dataset} & \textbf{\begin{tabular}[c]{@{}l@{}}Research entities \\ (June 2022)\end{tabular}} & \textbf{Sources} & \textbf{\begin{tabular}[c]{@{}l@{}}Change-tracking \&\\ provenance in RDF\end{tabular}} & \textbf{\begin{tabular}[c]{@{}l@{}}Disambiguation \\ method\end{tabular}} & \textbf{Internal ID} & \textbf{Access} & \textbf{License} \\ \hline

\begin{tabular}[c]{@{}l@{}}OpenAlex\end{tabular} & 240,000,000 & \begin{tabular}[c]{@{}l@{}}Crossref, Pubmed,\\ORCID, ROR, DOAJ,\\Unpaywall, arXiv,\\Zenodo, the ISSN International Centre,\\the Internet Archive's General Index\end{tabular} & - & \begin{tabular}[c]{@{}l@{}}PIDs \\ and heuristics\end{tabular} & \begin{tabular}[c]{@{}l@{}}+\end{tabular} & \begin{tabular}[c]{@{}l@{}}API,\\ web interface,\\ database snapshot\end{tabular} & CC0 \\ \hline

\begin{tabular}[c]{@{}l@{}}OpenAIRE \\ Research Graph\end{tabular} & 232,174,001 & \begin{tabular}[c]{@{}l@{}}ArXiv, PMC,\\ Software Heritage, \\ Zenodo, DOAJ, \\ OpenCitations, \\ Crossref, DataCite \\ and hundreds more\end{tabular} & - & \begin{tabular}[c]{@{}l@{}}PIDs \\ and heuristics\end{tabular} & \begin{tabular}[c]{@{}l@{}}+, \\ non persistent\end{tabular} & \begin{tabular}[c]{@{}l@{}}API, \\ web interface, \\ dump\end{tabular} & CC BY \\ \hline

\begin{tabular}[c]{@{}l@{}}Semantic Scholar\end{tabular} & 212,605,886 & \begin{tabular}[c]{@{}l@{}}Association for Computational Linguistics,\\ Cambridge University Press,\\ IEEE, PubMed, Springer Nature,\\ The MIT Press, Wiley,\\ arXiv, HAL, PubMed, and others\end{tabular} & - & \begin{tabular}[c]{@{}l@{}}PIDs via AI\end{tabular} & \begin{tabular}[c]{@{}l@{}}+\end{tabular} & \begin{tabular}[c]{@{}l@{}}API,\\ web interface,\\ dump\end{tabular} & OCD-By 1.0 \\ \hline

\begin{tabular}[c]{@{}l@{}}OpenCitations\\ Meta\end{tabular} & 98,243,101 & \begin{tabular}[c]{@{}l@{}}Crossref, DataCite,\\ \& NIH. \\ In future: \\ JaLC, OpenAIRE \\\& Dryad \end{tabular} & + & PIDs & + (OMID) & 

\begin{tabular}[c]{@{}l@{}}SPARQL, API, \\ dump\end{tabular} & CC0 \\ \hline
Wikidata & 39,864,447 & \begin{tabular}[c]{@{}l@{}}Wikimedia projects\\ \& Freebase\end{tabular} & +/-, not in standard RDF & PIDs & + & \begin{tabular}[c]{@{}l@{}}SPARQL, API, \\ web interface, \\ dump\end{tabular} & CC0 \\ \hline

\begin{tabular}[c]{@{}l@{}}Springer Nature \\ SciGraph\end{tabular} & 14,632,430 & \begin{tabular}[c]{@{}l@{}}Springer Nature\\ \& partners\end{tabular} & - & PIDs & - & \begin{tabular}[c]{@{}l@{}}web interface, \\ dump\end{tabular} & CC BY-NC \\ \hline
BioTea & 1,623,541 & PMC-OA & - & PIDs & - & dump & CC BY-NC \\ \hline

\begin{tabular}[c]{@{}l@{}}Open Research \\ Knowledge Graph\end{tabular} & 32,268 & \begin{tabular}[c]{@{}l@{}}Crossref,\\crowdsourcing\end{tabular} & +/-, not in RDF & \begin{tabular}[c]{@{}l@{}}PIDs\end{tabular} & \begin{tabular}[c]{@{}l@{}}+\end{tabular} & \begin{tabular}[c]{@{}l@{}}SPARQL, API, \\ web interface, \\ dump\end{tabular} & CC BY-SA \\ \hline

Scholarly Data & 5678 & \begin{tabular}[c]{@{}l@{}}Semantic Web\\ conferences and\\ workshops\end{tabular} & - & machine learning & + & \begin{tabular}[c]{@{}l@{}}SPARQL, API, \\ web interface, \\ dump\end{tabular} & CC BY
\end{tabular}%
}
\end{sidewaystable*}

\section{Methodology}\label{sec:methodology}

OpenCitations Meta is populated from input data in CSV format (i.e., tabular form). This choice is not accidental. We have found that data exposed by OpenCitations in CSV format (e.g., from COCI \parencite{opencitations_coci_2022}) are downloaded more frequently, in comparison to the same data in more structured formats (i.e., JSON Scholix and RDF N-Quads). This is due to the smaller file size (compared to N-Quads and Scholix) and, above all, to the higher readability of the tabular format for a human. The latter is the main reason why the input format adopted by OpenCitations Meta is CSV, to facilitate the future crowdsourcing of bibliographic metadata from human curatorial activities \parencite{heibi_crowdsourcing_2019}. 

The input table of OpenCitations Meta has 11 columns, corresponding to a linearization of the OCDM \parencite{daquino_opencitations_2020}: id, title, author, editor, publication date, venue, volume, issue, page, type, and publisher. For an in-depth description of how each field is structured, see \parencite{massari_how_2022}.

\begin{figure}[H]
    \centering
    \includegraphics[width=\textwidth]{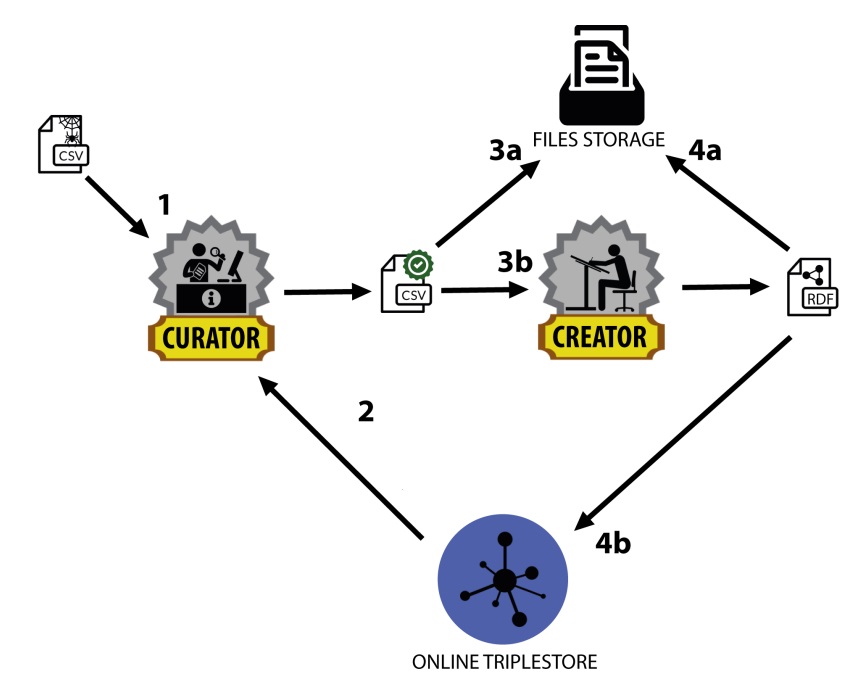}
    \caption{OpenCitations Meta workflow. First, the input data in CSV format is automatically corrected (1), deduplicated, and enriched with pre-existing information from within a triplestore (2). The corrected CSV is returned as output (3a). Second, the data are transformed into RDF (3b), saved to file (4a) and finally entered into the triplestore (4b).}
    \label{fig:meta_process}
\end{figure}

Once the CSV tabular data have been acquired, the data are first automatically curated (Curator step) and then converted to RDF based on the OCDM (Creator step). Finally, the curated CSV and RDF are stored as files, while a corresponding triplestore is incrementally populated. Fig. \ref{fig:meta_process} summarizes the workflow.

\subsection{Curator: deduplication, enrichment and correction}\label{sec:dedup}

The curation process performs three main actions to improve the quality of the received data: deduplication, enrichment, and correction. It is important to note that the entire curation process is fully automated, with no manual intervention involved.

The approach chosen for data deduplication is based strictly on identifiers. In other words, two different entities are considered the same if, and only if, both have the same identifier, e.g., a DOI for articles, an ORCID for people, an ISBN for books, and an ISSN for publication venues (e.g., journals). If multiple identifiers are available for an entity, all are taken into consideration in the deduplication process.

Different resources with the same identifier are merged following a precise rule: If the resources are part of the same CSV file, the information of the first occurrence is favored. However, if the resource is already described in the triplestore, the information in the triplestore will be favored. In other words, we consider the information stored in the triplestore as trusted, and  it can only be incremented with additional data coming from a CSV source.

Once an entity is deduplicated, it is assigned a new, permanent internal identifier called an OpenCitations Meta Identifier (OMID). The OMID has structure \\\texttt{[entity\_type\_abbreviation]/[supplier\_prefix][sequential\_number]}. For example, the first journal article ever processed has OMID \texttt{br/0601}, where \texttt{br} is the abbreviation of ``bibliographic resource'', and \texttt{060} corresponds to the supplier prefix, which indicates the database to which the bibliographic resource belongs (in this case,  OpenCitations Meta). Finally, \texttt{1} indicates that this OMID identifies the index's first bibliographic resource ever recorded for that prefix.

More precisely, the supplier prefix used for OpenCitations Meta is ``06[1-9]*0'', i.e., ``06'' optionally followed by any number excluding zero, and an ``0'' at the end. For example, ``060'', ``0610'', and ``06230'' are valid supplier prefixes in OpenCitations Meta. The reason for using a range of prefixes rather than a single one is strictly technical and internal. It's tied to our system's ability to execute operations in parallel. By having multiple prefixes, we can work on different directories simultaneously, each with a unique name corresponding to a supplier prefix. This approach optimizes our workflow and enhances efficiency. Note that, despite the variety in prefixes, all of them are under the umbrella of OpenCitations Meta.

The use of supplier prefixes in OMID identifiers serves multiple purposes. First, it allows for the identification of the source database for provenance tracking, which is crucial for automated metadata processing. The prefix ``060'' specifically indicates that OpenCitations Meta is the source of the metadata. Second, the structure of the prefix, ending with a ``0'', facilitates unambiguous automated parsing. The choice of ``06'' as the prefix for OpenCitations Meta is in line with the existing list of supplier prefixes used for Open Citation Identifiers (OCIs), which can be found in the OpenCitations OCI List \parencite{opencitations_opencitations_nodate}. This consistent use of supplier prefixes across different types of identifiers within the OpenCitations ecosystem enhances the robustness and traceability of the metadata.

The entities that are subject to deduplication and subsequently identified with an OMID are external identifiers (abbr. \texttt{id}), agent roles (i.e., authors, editors, publishers, abbr. \texttt{ar}), responsible agents (i.e., people and organizations, abbr. \texttt{ra}), resource embodiments (i.e., pages, abbr. \texttt{re}), and venues, volumes and issues (which are all bibliographic resources, abbr. \texttt{br}). Volumes and issues have OMIDs because they are treated as first-class citizens, not attributes of articles. This has the advantage of permitting one, for instance, to search for the papers within a specific issue, the volumes of a named journal, or journal issues published within a certain time period. In contrast, titles and dates are treated as literal values, not as entities.

\begin{sidewaysfigure}    
\includegraphics[width=\columnwidth]{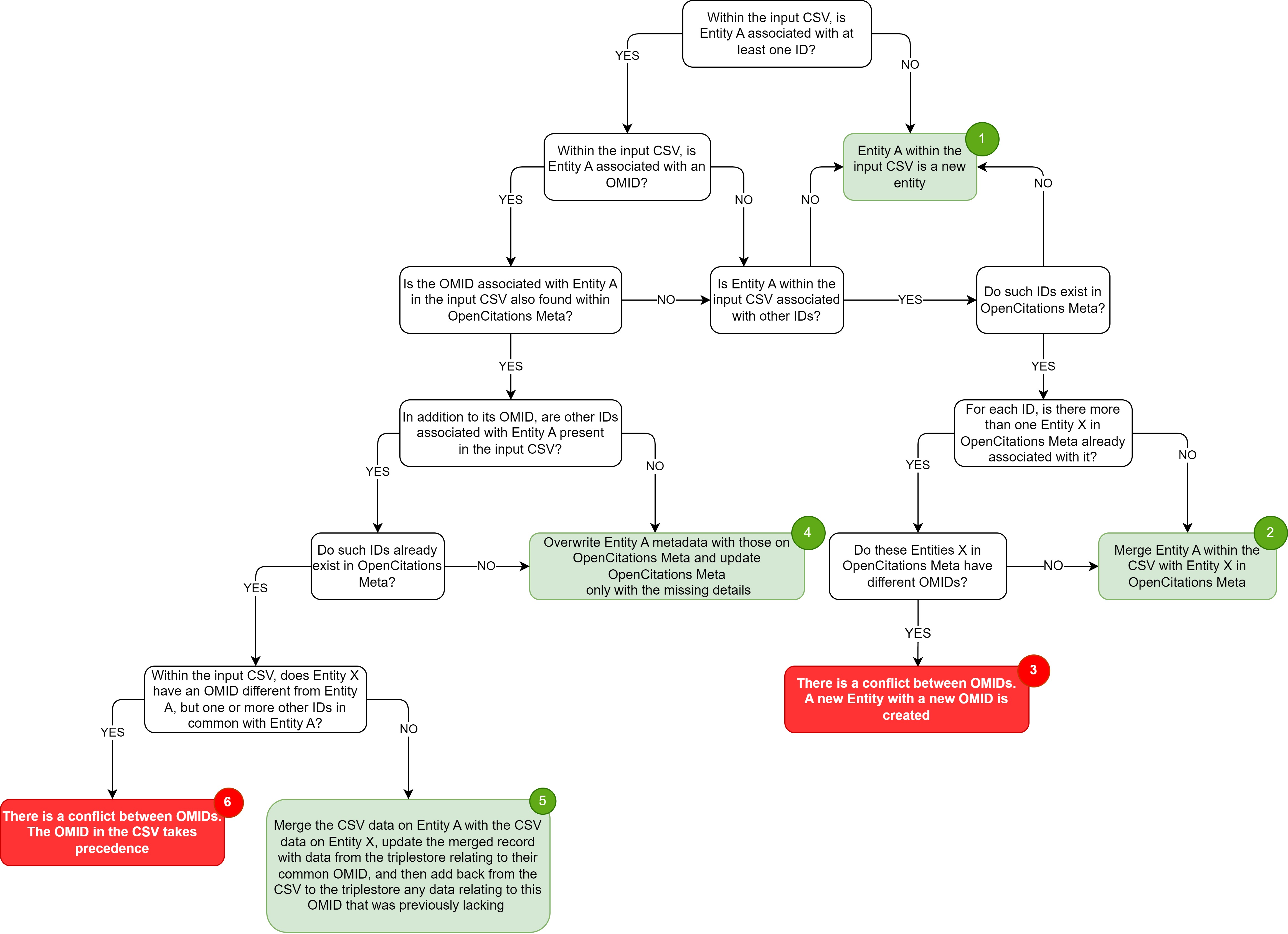}
    \caption{Deduplication decision tree.}
    \label{fig:dedup_tree}
\end{sidewaysfigure}

Fig. \ref{fig:dedup_tree} illustrates the deduplication decisional tree. Given an input entity and its identifiers, there are six possible outcomes:

\begin{enumerate}
    \item If the entity has no identifiers, or they do not exist in the triplestore, then a new OMID is created for the entity;
    \item If the entity does not have an OMID, and if one of its external identifiers has already been associated with one and only one other entity, then the two entities are merged and treated as the same;
    \item If the entity's external identifiers in the CSV connect two or more entities within the triplestore that had hitherto been distinct, and no OMID is specified in the CSV, then a conflict arises that cannot be resolved automatically and will require manual intervention. A new OMID is minted for this conflictual entity. For example, in the CSV, the same journal name is associated with two identifiers, issn:1588-2861 and issn:0138-9130; however, in the triplestore, there are entries for two separate entities, one with identifier issn:1588-2861 and the other with identifier issn:0138-9130, which in reality refer to the same entity;
    \item If an entity in the CSV has an OMID that exists in the triplestore and no other IDs are present, then the information in the triplestore overwrites that in the CSV. The triplestore is then updated only by the addition of missing details. In other words, specifying its OMID for an entity in the CSV is a way to update an existing entity within OpenCitations Meta; 
    \item If an entity has an existing OMID and additional identifiers are associated with other entities without an OMID (in the CSV) or with the same OMID (in the CSV or triplestore), then the entities are merged. Moreover, the information in the CSV is overwritten with that already available in the triplestore, and missing details present in the CSV are then added to the triplestore;
    \item Finally, if external identifiers connect several entities in the triplestore with different OMIDs, then a conflict arises. In this case, the OMID specified in the CSV takes precedence, and only entities with that OMID are merged.
\end{enumerate}

\begin{figure}[H]
    \centering
    \includegraphics[width=\textwidth]{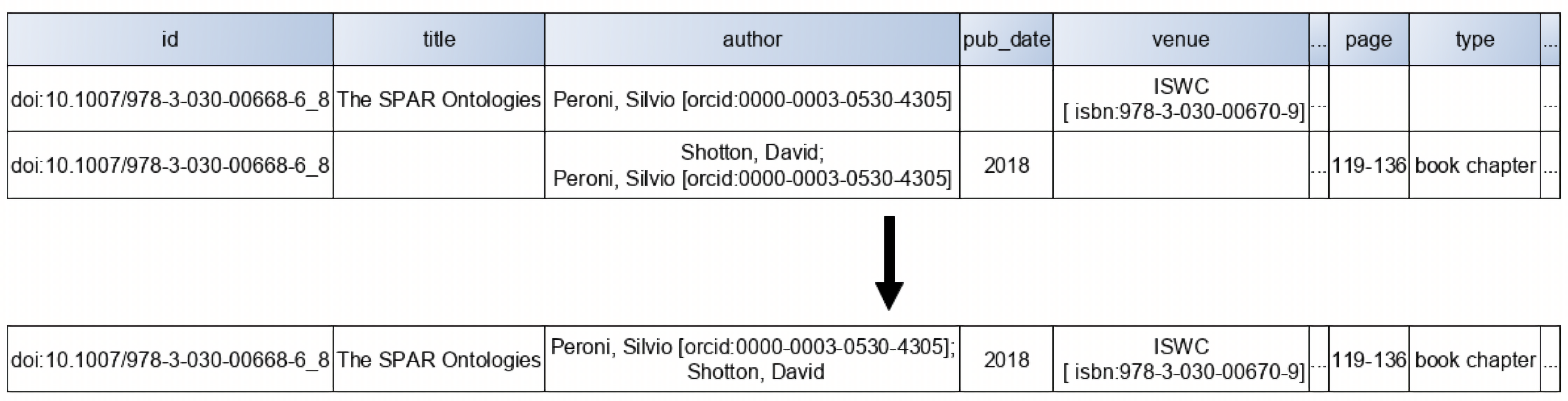}
    \caption{During a merge, the first information found takes precedence. In this example, David Shotton is inserted after Silvio Peroni in the list of authors because Peroni was already recorded as the first author, even if Shotton appears before Peroni in the second occurrence.}
    \label{fig:merge_order_ar}
\end{figure}

Given these general rules, three particular cases deserve special concern. The first notable issue concerns the order of authors and editors, which must be maintained according to the OCDM. In the event of a merge, the order recorded when the entity was first created overwrites subsequent ones, and any new authors or editors are added to the end of the existing list, as shown in Fig. \ref{fig:merge_order_ar}.

Second, in the context of two bibliographic resources being merged, the people involved as authors or editors without an identifier are disambiguated based on their given and family names.

The last significant case involves the containment relationship between articles, issues, volumes and venues. This structure is preserved in the case of a merge, where two volumes or issues are considered the same only if they have the same value, which may be a sequential number (e.g., ``Volume 1'') or an arbitrary name (e.g., ``Clin\_Sect'').

\subsection{Curator: error proofing}\label{sec:error_proofing}

Once all entities have obtained an OMID, data are normalized, and the errors that can be handled automatically are corrected. All identifiers are checked based on their identifier scheme -- for instance, the syntactic correctness of ISBNs, ISSNs and ORCIDs is computed using specific formulas provided by the documentation of the identifier scheme. However, the semantic correctness of identifiers is verified only for ORCIDs and DOIs, which is done using open APIs to verify their actual existence -- since, for instance, it is possible to produce an ORCID that is valid syntactically, but that is not in fact assigned to a person.

All ambiguous and alternative characters used for spaces (e.g., tab, no-break space, em space) are transformed into space (Unicode character U+0020). Similarly, ambiguous characters for hyphens within ids, pages, volumes, issues, authors and editors (e.g., non-breaking hyphens, en dash, minus sign) are changed to hyphen-minus (Unicode character U+002D).

Regarding titles of bibliographic resources (``venue'' and ``title'' columns), every word in the title is capitalized except for those with capitals within them (that are probably acronyms, e.g., ``FaBiO'' and ``CiTO''). This exception, however, does not cover the case of entirely capitalized titles. The same rule is also followed for authors and editors, whether individuals or organizations.

Dates are parsed considering both the format validity, based on ISO 8601 (YYYY-MM-DD) \parencite{wolf_date_1997}, and the value (e.g., 30 February is not a valid date). Where necessary, the date is truncated. For example, the date 2020-02-30 is transformed into 2020-02 because the day of the given date is invalid. Similarly, 2020-27-12 will be truncated to 2020 since the month (and hence the day) is invalid. The date is discarded if the year is invalid (e.g., a year greater than 9999).

The correction of volume and issue numbers is based on numerous rules which deserve special mention. In general, we have identified six classes of errors that may occur, and each different class is addressed accordingly:

\begin{enumerate}
\item Volume number and issue number in the same field (e.g., ``Vol. 35 N° spécial 1''). The two values are separated and assigned to the corresponding field.
\item Prefix errors (e.g., ``.38''). The prefix "." is deleted.
\item Suffix errors (e.g., ``19/''). The suffix "/" is deleted.
\item Encoding errors (e.g., ``5â\textbackslash x80\textbackslash x926'', ``38â39'', ``3???4''). Only numbers at the extremes are retained, separated by a single hyphen. Therefore, the examples are corrected to ``5-6'', ``38-39'', and ``3-4'', respectively, since ``â\textbackslash x80\textbackslash x92'', ``â'' and ``???'' are incorrectly encoded hyphens.
\item Volume classified as issue (e.g., ``Volume 1'' in the ``issue'' field). If the volume pattern is found in the ``issue'' field and the ``volume'' field is empty, the content is moved to the ``volume'' field, and the ``issue'' field is set to null. However, if the ``issue'' field contains a volume pattern and the ``volume'' field contains an issue pattern, the two values are swapped.
\item Issue classified as volume (e.g., ``Special Issue 2'' in the ``volume'' field). It is handled in the same way as case 5, but in reversed roles.
\end{enumerate}

We considered as volumes those patterns containing the words ``original series'', ``volume'', ``vol'', and volume in various other languages, e.g., ``tome'' in French and ``cilt'' in Turkish. For example, ``Original Series'', ``Volume 1'', ``Vol 71'', ``Tome 1'', and ``Cilt: 1'' are classified as volumes. Similarly, we considered as issues those patterns containing the words ``issue'', ``special issue'' and issue in various languages, e.g., ``hors-série'' (special issue in French) and ``özel sayı'' (special issue in Turkish). For example, ``issue 2'', ``special issue 2'', ``Special issue 'Urban Morphology''', ``Özel Sayı 5'', and ``Hors-série 5'' are classified as issues.

Finally, where a value is both invalid in its format and invalid because it is in the wrong field, then such a value is first corrected and then moved to the right field, if appropriate.

Once the input data has been disambiguated, enriched and corrected, a new CSV file is produced and stored. This file represents the first output of the process (3a in Fig. \ref{fig:meta_process}).

\subsection{Creator: semantic mapping}\label{subsec:semantic_mapping}

\begin{figure}[H]
    \centering
    \includegraphics[width=\textwidth]{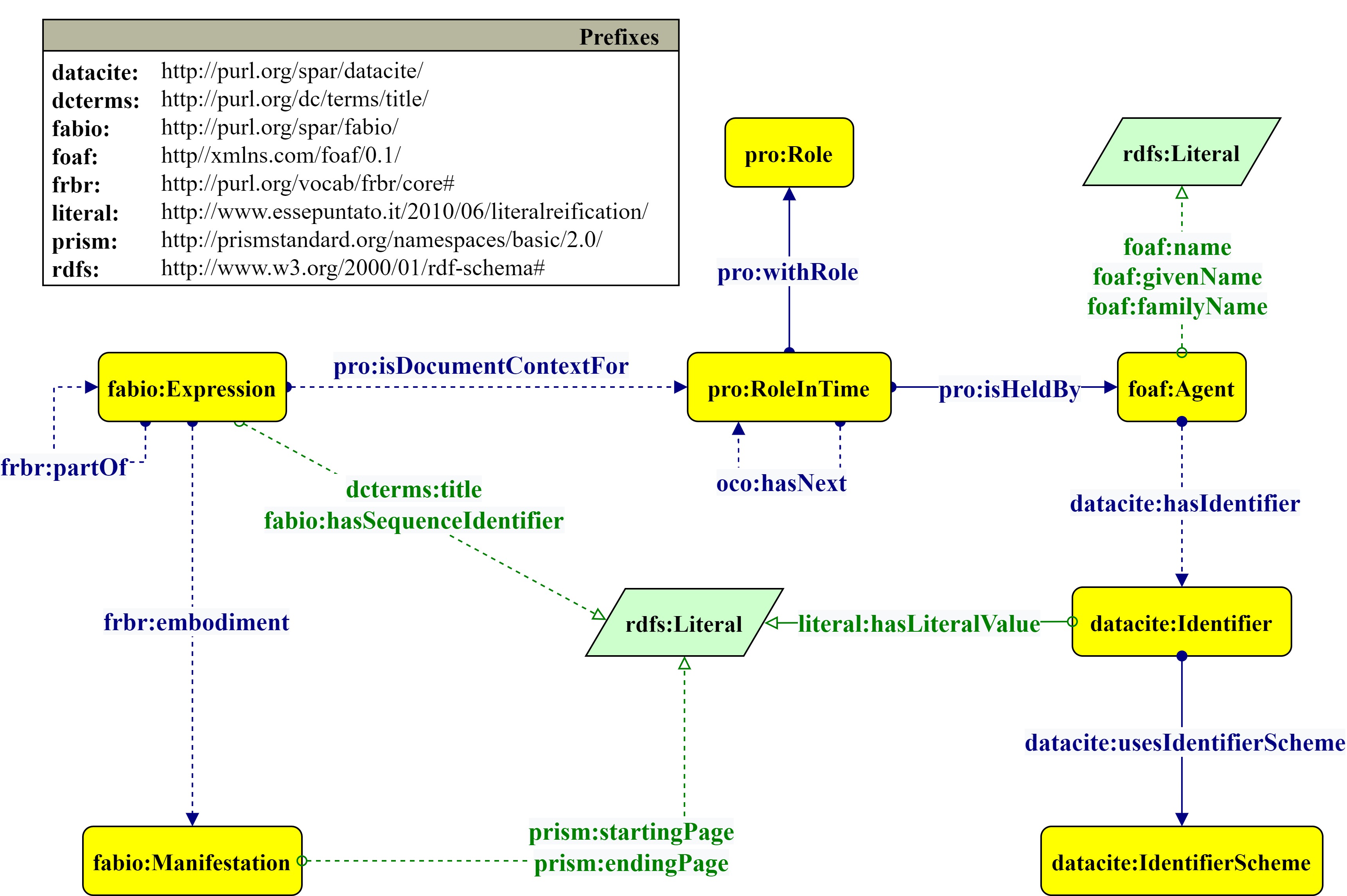}
    \caption{Part of the OCDM used in OpenCitations Meta. Yellow rectangles represent classes, green polygons represent datatypes, and blue and green arrows represent object properties and data properties, respectively.}
    \label{fig:data_model}
\end{figure}

In this phase, data are modeled in RDF following the OCDM \parencite{daquino_opencitations_2020}. This ontology reuses entities defined in the SPAR Ontologies to represent bibliographic entities (\texttt{fabio:Expression}), identifiers (\texttt{datacite:Identifier}), agent roles (\texttt{pro:RoleInTime}), responsible agents (\texttt{foaf:Agent}) and publication format details\\ (\texttt{fabio:Manifestation}). The agent role (i.e., author, editor or publisher) is used as a proxy between the bibliographic resource and the responsible agent, i.e., the person or organization. This approach helps us define time-dependent and context-dependent roles and statuses, such as the order of the authors \parencite{peroni_scholarly_2012}. Fig. \ref{fig:data_model} depicts the relationships between the various entities through the Graffoo graphical framework \parencite{presutti_modelling_2014}. 

For example, in OpenCitations Meta the entity with OMID \texttt{omid:br/062601067530} has title \emph{Open Access And Online Publishing: A New Frontier In Nursing?} (\texttt{dcterms:title}), and was published on 2012-07-25 (\texttt{prism:publicationDate}). Using FRBR \parencite{tillett_what_2005}, the article is the final published version, or an expression of the original work (\texttt{fabio:Expression}), which has as a sample the entity \texttt{omid:re/06260837633} (\texttt{frbr:embodiment}), that is the printed publication corresponding to pages 1905-1908 of the journal volume (\texttt{prism:startingPage}, \texttt{prism:endingPage}). More precisely, the article is part of (\texttt{frbr:partOf}) the issue (\texttt{fabio:JournalIssue}) number 9\\ (\texttt{fabio:hasSequenceIdentifier}), contained in the volume (\texttt{fabio:JournalVolume}) number 68 of the venue \emph{Journal Of Advanced Nursing} (\texttt{fabio:Journal}).

Furthermore, the person (\texttt{foaf:Agent}) Glenn Hunt (\texttt{foaf:givenName},\\ \texttt{foaf:familyName}) is the first author (\texttt{pro:RoleInTime}) in the context of this article \\(\texttt{pro:isDocumentContextFor}). Similarly, the second author is Michelle Cleary\\ (\texttt{pro:hasNext}). 

Finally, this publication has the OpenCitations Meta Identifier (OMID)\\ \texttt{omid:id/062601093630} \\(\texttt{datacite:hasIdentifier}), an entity of type \texttt{datacite:Identifier}. It also has an external identifier,  that uses as its identifier scheme a Digital Object Identifier (DOI) (\texttt{datacite:usesIdentifierScheme}) and that has the literal value ``10.1111/j.1365-2648.2012.06023.x'' (\texttt{literal:hasLiteralValue}).

Once the mapping is complete, the RDF data produced can be stored (4a in Fig. \ref{fig:meta_process}) and uploaded to a triplestore (4b in Fig. \ref{fig:meta_process}).

\subsection{Creator: provenance and change tracking}\label{subsec:provenance}
In addition to handling their metadata, great importance is given to provenance and change tracking for entities in OpenCitations Meta. Provenance is a record of who processed a specific entity by creating, deleting, modifying or merging it, when this action was performed, and what the primary source was \parencite{gil_provenance_2010}. Keeping track of this information is crucial to ensure the reliability of the metadata within OpenCitations Meta. Indeed, the truth of a statement on the Web and the Semantic Web is never absolute, and integrity must be assessed by every application that processes information by evaluating its context \parencite{koivunen_semantic_2001}.

However, besides storing provenance information, mechanisms to understand the evolution of entities are critical when dealing with activities such as research assessment exercises, where modifications, due to either corrections or mis-specification, may affect the overall evaluation of a scholar, a research group, or an entire institution. For instance, the name of an institution might change over time, and the reflection of these changes in a database ``make[s] it difficult to identify all institution's names and units without any knowledge of institution's history'' \parencite{pranckute_web_2021}. This scenario can be prevented by keeping track of how data evolved in the database, thus enabling users to understand such dynamics without accessing external background knowledge. To our knowledge, no other semantic database of scholarly metadata keeps track of changes and provenance in standard RDF 1.1.

\begin{figure}[H]
    \centering
    \includegraphics[width=\textwidth]{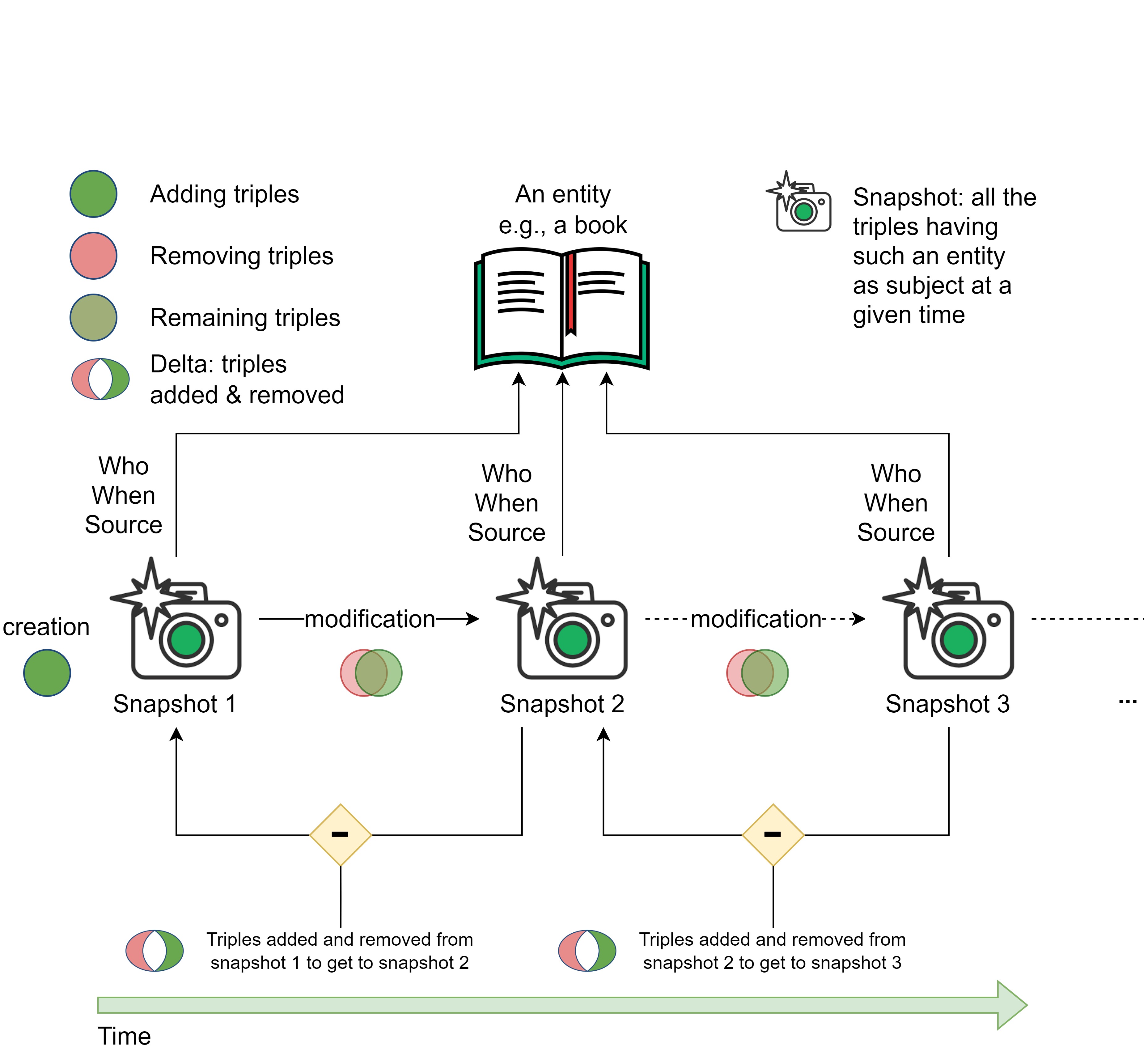}
    \caption{A high-level description of the provenance layer of the OCDM to keep track of the changes to an entity. To keep track of the full history of an entity, we need to store all the triples of its most recent snapshot plus all the deltas built by modifying the previous snapshots.}
    \label{fig:snapshots}
\end{figure}

The provenance mechanism employed by OpenCitations describes an initial creation snapshot for each stored entity, potentially followed by other snapshots detailing modification, merge or deletion of data, each marked with its snapshot number, as summarised in Fig. \ref{fig:snapshots}

Regarding the semantic representation, the problem of provenance modeling \parencite{sikos_provenance-aware_2020} and change-tracking in RDF \parencite{pelgrin_towards_2021} has been discussed in the scholarly literature. To date, no shared standard achieves both purposes. For this reason, OpenCitations employs the most widely shared approaches, i.e., named graphs \parencite{carroll_named_2005}, the Provenance Ontology \parencite{lebo_prov-o_2013}, and Dublin Core \parencite{dcmi_usage_board_dcmi_2020}.

In particular, each snapshot is connected to the previous one via the \texttt{prov:wasDerivedFrom} predicate and is linked to the entity it describes via \texttt{prov:specializationOf}. In addition, each snapshot corresponds to a named graph in which the provenance metadata are described, namely the responsible agent (\texttt{prov:wasAttributedTo}), the primary source (\texttt{prov:hadPrimarySource}), the generation time (\texttt{prov:generatedAtTime}), and, after the generation of an additional snapshot, the invalidation time (\texttt{prov:invalidatedAtTime}). Each snapshot may also optionally be represented by a natural language description of what happened (\texttt{dcterms:description}).

\begin{figure}[H]
    \centering
    \includegraphics[width=\textwidth]{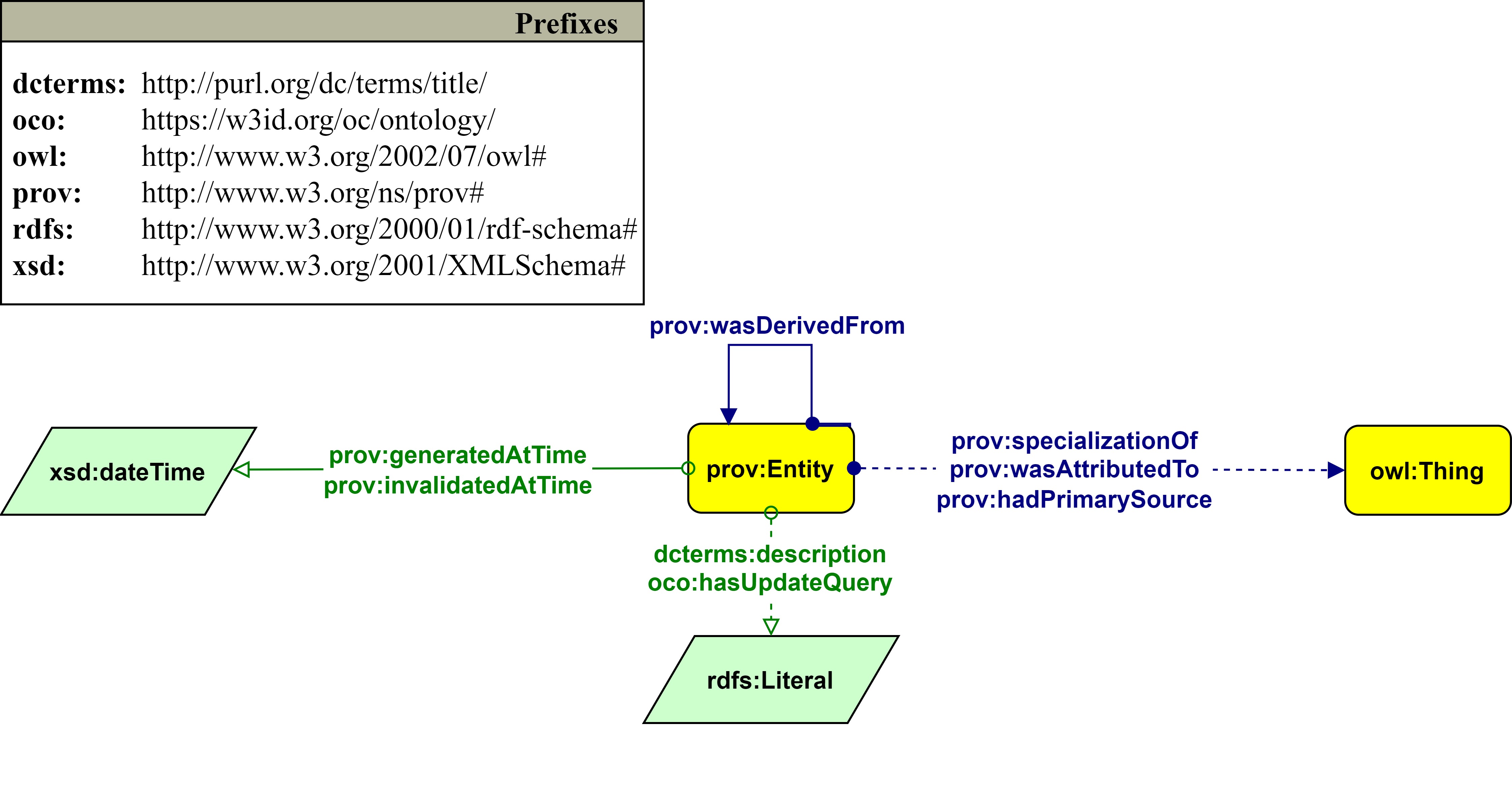}
    \caption{The Graffoo diagram describing snapshots (\texttt{prov:Entity}) of an entity (linked
via \texttt{prov:specializationOf}) and the related provenance information.}
    \label{fig:ocdm_prov}
\end{figure}

In addition, the OCDM provenance model adds a new predicate, \texttt{oco:hasUpdateQuery}, described within the OpenCitations Ontology \parencite{daquino_oco_2019}, which expresses the delta between two versions of an entity via a SPARQL UPDATE query. Fig. \ref{fig:ocdm_prov} displays the model via a Graffoo diagram.

The deduplication process described in Section \ref{sec:dedup} takes place not only on the current state of the data set but on its entire history by enforcing the change-tracking mechanism. In other words, if an identifier can be traced back to an entity deleted from the triplestore, that identifier will be associated with the OMID of the deleted entity. If the deletion is due to a merge chain, the OMID of the resulting entity takes precedence. For more on the time-traversal queries methodology, see \parencite{massari_performing_2022}. For more details on the programming interface for creating data and tracking changes according to the SPAR Ontologies, consult \parencite{groth_programming_2022}.

\section{Data and services}\label{sec:eval}

At the time of its initial release in December 2022, OpenCitations Meta included Crossref \parencite{hendricks_crossref_2020}, DataCite \parencite{brase_datacite_2010}, and the NIH Open Citation Collection (containing data from PubMed)  \parencite{icite_icite_2022} as its primary sources for the bibliographic metadata describing the publications  involved in citations within the following OpenCitations Indexes: COCI (\url{https://opencitations.net/index/coci}) \parencite{opencitations_coci_2022}, DOCI (\url{https://opencitations.net/index/doci}), and POCI (\url{https://opencitations.net/index/poci}). From a quantitative point of view, there are within this initial release of OpenCitations Meta 98,243,101 bibliographic entities (\texttt{fabio:Expression}), 309,881,223 authors (\texttt{pro:author}), 2,406,510 editors (\texttt{pro:editor}), 19,076 publishers (\texttt{pro:publisher}), and 659,214 venues\\ (e.g., resources of type \texttt{fabio:AcademicProceedings}, \texttt{fabio:ExpressionCollection}, \texttt{fabio:Book}, \texttt{fabio:BookSeries}, \texttt{fabio:Journal}, \texttt{fabio:ReferenceBook}, or \texttt{fabio:Series}) (Table \ref{tab:quantitative-overview}). Thus, on average, each bibliographic resource has three authors. Typically no editor is recorded, as the latter metadata are little used in our sources. In total, the triplestore consists of 3,749,729,755 triples (excluding provenance).

\begin{table}[H]
\centering
\caption{Quantitative overview of OpenCitations Meta at initial release.}
\label{tab:quantitative-overview}
\begin{tabular}{ll}
\textbf{Entity Type} & \textbf{Count} \\
\hline
Bibliographic Entities & 98,243,101 \\
Authors & 309,881,223 \\
Editors & 2,406,510 \\
Publishers & 19,076 \\
Venues & 659,214 \\
\end{tabular}
\end{table}

Editors and authors have been counted as roles, without disambiguating the individuals holding these roles. Conversely, bibliographic entities, publishers, and venues were counted by OMID. However, for venues (e.g., journals), we have taken an extra precaution: many are duplicated in OpenCitations Meta because they have no identifiers other than the OMID. Therefore, in the figures shown above, we found it reasonable to disambiguate the venues by title in the absence of other identifiers.

As shown in Table \ref{tab:top_pub_by_ven}, Springer Science is the publishing entity with the highest number of venues (2097), followed by Elsevier BV (1961) and IEEE (1775). When counting the number of publications, Elsevier is in the lead (16,933,610), followed by Springer Science (11,507,498) and Wiley (7,262,893), as shown in Table \ref{tab:top_pub_by_pub}.

Considering the venues in Table \ref{tab:top_ven_by_pub}, Wiley's ChemInform has the most publications (421,735), followed by Elsevier's SSRN Electronic Journal (337,223) and Springer's Journal On Data Semantics (330,093).

Table \ref{tab:top_type_by_pub} lists all the types of bibliographic resources in OpenCitations Meta. The current data set contains mostly journal articles (67,904,323), which exceed the number of book chapters in second place (6,476,623) by about 10 times, and proceedings articles in third place (5,046,165) by about 13 times.

Figure \ref{fig:top_year_by_pub}, which lists the number of publications per year, shows an increasing trend, with a greater number of publications from year to year.

\begin{table}[H]
\caption{The top 10 publishers by number of venues.}
\label{tab:top_pub_by_ven}
\resizebox{\textwidth}{!}{
\begin{tabular}{ll}
\textbf{Publisher}                                        & \textbf{Number of venues}  \\ 
\hline
Springer Science and Business Media LLC                   & 2097                    \\
Elsevier BV                                               & 1961                    \\
Institute of Electrical and Electronics Engineers (IEEE)  & 1775                  \\
Wiley                                                     & 1020                    \\
Informa UK Limited                                        & 934                    \\
SAGE Publications                                         & 362
        \\
Oxford University Press (OUP)                             & 260                    \\
Cambridge University Press (CUP)                          & 235                    \\
Association for Computing Machinery (ACM)                 & 194
       \\
Ovid Technologies                                         & 170             
\end{tabular}
}
\end{table}

\begin{table}[H]
\caption{The top 10 publishers by number of publications.}
\label{tab:top_pub_by_pub}
\resizebox{\textwidth}{!}{
\begin{tabular}{ll}
\textbf{Publisher}                                       & \textbf{Number of publications} \\ \hline
Elsevier BV                                              & 16,933,610                      \\
Springer Science and Business Media LLC                  & 11,507,498                      \\
Wiley                                                    & 7,262,893                      \\
Institute of Electrical and Electronics Engineers (IEEE) & 4,440,816                       \\
Informa Uk Limited                                       & 4,026,383                       \\
Oxford University Press (Oup)                            & 1,806,341                       \\
Sage Publications                                        & 1,758,588                       \\
Ovid Technologies (Wolters Kluwer Health)                & 1,515,620                       \\
American Chemical Society (ACS)                          & 1,461,570                      \\
The Global Biodiversity Information Facility             & 1,071,159                      
\end{tabular}
}
\end{table}

\begin{table}[H]
\caption{The top 10 venues by number of publications.}
\label{tab:top_ven_by_pub}
\resizebox{\textwidth}{!}{
\begin{tabular}{ll}
\textbf{Venue}                           & \textbf{Number of publications} \\ \hline
ChemInform                               & 421,735                \\
SSRN Electronic Journal                  & 337,223                \\
Lecture Notes in Computer Science        & 330,093                \\
The Lancet                               & 288,366                \\
PLOS ONE                                 & 273,682                \\
Journal of Biological Chemistry          & 267,669                \\
Chemischer Informationsdienst            & 239,077                \\
Nature                                   & 210,842                \\
Journal of the American Chemical Society & 190,066                \\
SPIE Proceedings                         & 184,280               
\end{tabular}
}
\end{table}

\begin{table}[H]
\caption{All the bibliographic resource types involved in OpenCitations Meta, sorted by the number of publications of that type. The reference ontologies are FaBiO (\url{http://purl.org/spar/fabio}), DOCO (\url{http://purl.org/spar/doco}), and FAIR reviews (\url{http://purl.org/spar/fr}).}
\label{tab:top_type_by_pub}
\resizebox{\textwidth}{!}{
\begin{tabular}{ll}
\textbf{Bibliographic resource type}         & \textbf{Number of publications} \\ \hline
Journal article (\texttt{fabio:JournalArticle})       & 67,904,323                     \\
Book chapter (\texttt{fabio:BookChapter})             & 6,476,623                      \\
Proceedings article (\texttt{fabio:ProceedingsPaper}) & 5,046,165                     \\
Journal issue (\texttt{fabio:JournalIssue})           & 4,862,169                        \\
Book (\texttt{fabio:Book})                            & 3,254,002                        \\
Journal volume (\texttt{fabio:JournalVolume})         & 1,469,210                      \\
Dataset (\texttt{fabio:DataFile})                      & 1,434,704                         \\
Web content (\texttt{fabio:WebContent})               & 334,015                         \\
Report (\texttt{fabio:ReportDocument})                 & 293,827                    \\
Reference book (\texttt{fabio:ReferenceBook})         & 204,667                          \\
Reference entry (\texttt{fabio:ReferenceEntry})       & 186,832                         \\
Journal (\texttt{fabio:Journal})                      & 187,722                       \\
Proceedings (\texttt{fabio:AcademicProceedings})     & 68,961                           \\
Book series (\texttt{fabio:BookSeries})                & 32,948                           \\
Dissertation (\texttt{fabio:Thesis})                  & 21,974                         \\
Standard (\texttt{fabio:SpecificationDocument})       & 13,797                         \\
Book section (\texttt{fabio:ExpressionCollection})    & 9227                  \\
Series (\texttt{fabio:Series})                        & 8967                            \\
Peer review (\texttt{fr:ReviewVersion})               & 4677                            \\
Book part (\texttt{doco:Part})                        & 2544                            \\
Book set (\texttt{fabio:BookSet})                     & 47                             
\end{tabular}
}
\end{table}

\begin{figure}[H]
    \centering
    \caption{Top 10 years of publication by the number of publications in that year.}
    \label{fig:top_year_by_pub}
    \begin{tikzpicture}
        \begin{axis}[
            width=\textwidth,
            ybar,
            enlargelimits=0.15,
            legend style={at={(0.5,-0.15)},
            anchor=north,legend columns=-1},
            ylabel={Number of publications},
            symbolic x coords={2013, 2014, 2015, 2016, 2017, 2018, 2019, 2020, 2021, 2022 (incomplete)},
            xtick=data,
            nodes near coords,
            nodes near coords align={vertical},
            x tick label style={rotate=45,anchor=east},
            ]
            \addplot coordinates {(2013, 2720717) (2014, 2861672) (2015, 3011018) (2016, 3216676) (2017, 3359064) (2018, 3549528) (2019, 3736978) (2020, 4143432) (2021, 4186834) (2022 (incomplete), 3346850)};
        \end{axis}
    \end{tikzpicture}
\end{figure}

OpenCitations Meta allows the users to explore such data either via SPARQL (\url{https://opencitations.net/meta/sparql}) or via an API (\url{https://opencitations.net/meta/api/v1}). In particular, the OpenCitations Meta API retrieves a list of bibliographic resources and related metadata starting from one or more publication identifiers, an author's ORCID, or an editor's ORCID. Textual searches are currently under testing and will be released in the future as one further operation of the OpenCitations Meta API. In particular, text searches on titles, authors, editors, publishers, IDs, and venues can be performed. They can also be achieved on volume and issue numbers, provided the venue is first specified. Indeed, searches on multiple fields can be combined using the Boolean conjunction and disjunction operators. For example, once the operation is released, the user will be able to search for all bibliographic resources whose title contains the word ``micro-chaos'' published either by Philosophical Studies or the Journal of Nonlinear Science: \emph{title=micro-chaos\&\&venue=philosophical\%20studies||title=micro-chaos\&\&venue=journal\%20of\%20nonlinear\%20science}, where ``\&\&'' is the conjunction operator, while || is the disjunction operator.

Finally, all data and provenance are available as dumps in RDF (JSON-LD) \parencite{opencitations_opencitations_2023-1} or CSV format \parencite{opencitations_opencitations_2023} under a CC0 license.

\section{Discussion}\label{sec:discussion}

This section aims to address key decision points and trade-offs involved in the development of OpenCitations Meta, particularly focusing on its advantages and disadvantages compared to other bibliographic data sets.

The inception of OpenCitations Meta was driven by both internal operational challenges and broader community needs. Relying on external APIs introduced network latency, potential rate limits, and reliability concerns, leading to performance bottlenecks in our citation indexes. Moreover, this external dependency limited our textual search capabilities. By transitioning to an internal database, we aimed to mitigate these issues and optimize our system's efficiency.

Furthermore, OpenCitations Meta plays a pivotal role in achieving the deduplication of citations within the OpenCitation Index. Within Meta, diverse entities coming from different data sources can be merged into unified entities. This fact occurs when two distinct entities share a common, defined PID (e.g., DOI). Citations are generated based on the newly merged entities stored in OpenCitations Meta, referred to as OMID to OMID citations.

From a community standpoint, OpenCitations Meta functions as an aggregator rather than an isolated database. It aims to consolidate existing sources of open bibliographic metadata into a unified framework. We advocate for a bibliographic landscape that operates as a network of interconnected infrastructures, each addressing unique challenges.

One of the pivotal decisions in the development of OpenCitations Meta was the introduction of new identifiers (OMIDs). This was primarily aimed at managing provenance and change tracking effectively. When a resource is modified, it remains the same entity, and having a consistent identifier for it is essential for maintaining a coherent and traceable record. Additionally, not all resources have a persistent external identifier, making it necessary to assign new IDs for internal consistency and external reference.

We acknowledge that introducing new identifiers can add complexity to the identifier ecosystem. However, new IDs do not inherently cause misalignment. Automatic correction policies can be implemented, for example, using update data from Crossref. Although we currently do not employ such mechanisms due to implementation complexities, it is a feasible avenue for future work. Human curation would, of course, take precedence in any such system.

We are also exploring various initiatives that align with this approach, such as the COAR Notify Initiative \parencite{klein_notify_2021}, and SPARQL notifications (SEPA) \parencite{roffia_sparql_2020}. These technologies could potentially allow for better integration and updates.

From a broader perspective, we share the sentiment about promoting more reuse and less recreation. The introduction of OpenCitations Meta was deemed necessary because there was no existing solution that met our specific needs for provenance, change tracking, and the ability to handle resources without persistent external identifiers.

OpenCitations Meta has several advantages, including better integration with existing data ecosystems and a fully semantic-web approach. However, it also has its limitations, such as the added complexity and the challenges of integrating with other data sets. These trade-offs are justified given the benefits in terms of provenance tracking and the potential for future integration with systems that could allow for automatic updates and corrections.

As shown in Section \ref{sec:related_works}, when considering only data sets represented using Semantic Web technologies, OpenCitations Meta, which currently includes data from Crossref, DataCite, and PubMed (via the NIH Open Citation Collection) \parencite{icite_icite_2022}, is the largest in terms of data volume. Plans are underway to ingest data from additional sources, including the Japan Link Center \parencite{hara_introduction_2020}, the OpenAIRE Research Graph \parencite{grana_openaire_2017}, and the Dryad Digital Repository \parencite{vision_dryad_2010}. Moreover, we are actively working on the alignment of OMIDs with OpenAlex IDs \parencite{priem_openalex_2022}.

In contrast to the OpenAIRE Research Graph, which utilizes internal identifiers that are not persistent, OpenCitations Meta offers enhanced functionality through the use of OMIDs—globally unique persistent identifiers that can be used for provenance and change-tracking of every entity. Moreover, this usage makes it possible to represent and index citations between bibliographic resources that lack an external persistent identifier such as a Digital Object Identifier (DOI). This feature adds significant value for the OpenCitations Indexes, as it allows for the first time the ingestion of many citations which until now were not possible to be characterized, particularly citations between publications from the Humanities and Social Sciences \parencite{gorraiz_availability_2016}, and citations involving primary sources, e.g., a statue, a painting, or a codex, which typically lack a persistent identifier. Importantly, having an OMID also permits the identified resource to be assigned a unique and persistent URL, for example \url{https://w3id.org/oc/meta/br/061401975837} for omid:br/061401975837.

Another feature that, to the best of our knowledge, is only present in OpenCitations Meta is the mechanism for change-tracking management within the provenance information stored in RDF. This information can be queried using the Python time-agnostic-library software \parencite{massari_performing_2022}. It can perform time-traversal SPARQL queries, i.e., queries across different snapshots together with provenance information.

As far as other bibliographic data sets that do not use Semantic Web technologies go, OpenAlex \parencite{priem_openalex_2022} is an important case to consider for comparison with OpenCitations Meta. OpenAlex uses web crawls to add missing metadata, a feature that allows it to automatically correct a higher number of errors appearing in the data of the sources, when compared to OpenCitations Meta.

Indeed, currently, the main limitation of OpenCitations Meta concerns the quality of the data, which is strictly dependent on the quality of information within its sources. Crossref does not double-check the metadata provided by publishers, and thus many errors are preserved. For instance, it is possible to encounter articles published in the future (the metadata available at \url{https://api.crossref.org/v1/works/10.12960/tsh.2020.0006} say that the article will be published in print in 2029). Some of these errors can be corrected automatically without any background knowledge, but others require either the use of web crawlers or manual intervention. Although OpenAlex is pursuing the path of web crawls, OpenCitations is working to develop a framework that will allow the editing and curation of data by trusted human domain experts (such as academic librarians).

OpenCitations Meta fulfills its primary purpose by holding the bibliographic metadata required to describe the citing and cited publications involved in the citations within the OpenCitations Indexes. In addition to these bibliographic metadata elements, however, we are well aware that there are additional metadata elements of great importance for the academic community: Abstracts, for text mining, domain and subject field determination, and indexing (even if the full texts of the publications are available open access elsewhere), and Funder IDs, Funding information and Institutional identifiers, essential for determining performance metrics and undertaking research assessment. Once we have completed the provision of our textual search operations, expanded our coverage in the ways indicated, and enhanced the computational infrastructure upon which OpenCitations Meta and the OpenCitations Indexes run, we will proceed to integrate and populate these additional metadata fields.

The provision of high-quality bibliographic metadata is a complex and difficult goal to achieve by automated operations, and the scale of the operations precludes manual curation except for a minority of records. No bibliographic data set is currently able to achieve this goal on its own. For this reason, all the available bibliographic databases should be viewed as complementary. For example, although at the moment OpenAlex provides more complete metadata, OpenCitations Meta has complete provenance data openly available, and enables more complex searches, thanks to the potentialities given by Semantic Web technologies. For example, "Search for all authors who coauthored with Silvio Peroni or Fabio Vitali in conference proceedings that were published by Springer after 2009". Furthermore, OpenAlex is only partially free, as a fee must be paid to make more than 100,000 requests per day via the API and to access data updated every hour via the API (instead of every month via the dump)\footnote{https://openalex.org/pricing}. In contrast, users can make unlimited requests to the latest version of OpenCitations Meta for free. 

Also, although the OpenAIRE Research Graph currently contains more metadata, such data are released under a CC-BY attribution license, whereas the data released by OpenCitations Meta are under a CC0 public domain waiver, permitting complete freedom for reuse, including commercial reuse, and for machine processing without any requirement for attribution. 

\section{Conclusion}\label{sec:conclusion}

This article detailed the methodology used to develop OpenCitations Meta, a database that stores and delivers bibliographic metadata for all publications involved in the citations documented in the OpenCitations Indexes. This process involves two main phases: (1) an automatic curation analysis aimed at deduplicating entities, correcting errors and enriching information, and (2) a data conversion to RDF, while keeping track of changes and provenance in RDF. 

Information about new publications is continually being added to Crossref, DataCite, and PubMed, and we will develop procedures to ingest these new metadata into OpenCitations Meta in a regular and timely manner. Furthermore, work is already underway to ingest bibliographic metadata from other sources as our human and computational resources permit. All ingested data will be openly published under a CC0 public domain waiver, in line with OpenCitations' commitment to open data.

OpenCitations Meta offers a plethora of advantages to the scholarly community. Foremost, its provenance model, grounded in the PROV Ontology, ensures transparency and traceability. Every bibliographic resource is enriched with detailed provenance information, allowing users to trace the origins and subsequent modifications of any data, fostering trust in the data's authenticity. 

Additionally, the change-tracking mechanism ensures the accuracy and reliability of data. Every modification, be it an addition, deletion, or merge, is recorded, granting users a granular view of resource evolution, a feature unparalleled in other bibliographic indexes.

Furthermore, OpenCitations Meta's commitment to semantic data ensures superior interoperability, aligning with the FAIR principles. The graph-based nature of RDF allows for a detailed web of data, where entities and their relationships are clearly defined and interconnected, ensuring external systems can comprehend not just the data, but also the context and relationships surrounding it.

From an infrastructural perspective, OpenCitations Meta introduces three pivotal advancements. First, OpenCitations Meta speeds search operations to retrieve metadata on publications involved in the citations stored in the OpenCitations Citation Indexes, since these metadata are now kept in-house, rather than being retrieved by on-the-fly API calls to external resources. Second, the use of OMIDs (OpenCitation Meta Identifiers) for all stored entities enables OpenCitations Meta to act as a mapping and disambiguation hub for publications that may have more than one external PID (for example a journal article described in Crossref with a DOI (Digital Object Identifier), and the same publication described in PubMed with a PMID (PubMed Identifier), while also making it possible to characterise citations involving resources lacking any external PIDs. Consequently, the third benefit is that OpenCitations Meta will allow citations defined as OMID-to-OMID to be described in a new unified OpenCitations Index, that will incorporate and disambiguate citations between documents currently using different identifier schemes and present in separate OpenCitations Indexes, e.g., in COCI, represented as DOI-to-DOI citations in Crossref, and in POCI, represented as PMID-to-PMID citations in PubMed. 

Future challenges will be to elaborate a disambiguation system for people lacking an ORCID identifier, to improve the quality of the existing metadata, to enhance the search operations and the storage efficiency, to add additional metadata fields for Abstracts, Funder IDs, Funding information, and Institutional identifiers, and to populate these where these metadata are available from our sources. 

Finally, an interface will be implemented and made available to trusted domain experts to permit direct real-time manual curation of metadata held by OpenCitations Meta. Such a system will track changes and provenance, will preserve the delta between different versions of each entity, and will retain information such as the agent responsible for the change, the primary source, and the date. In this way, we will strive to make OpenCitations Meta not only comprehensive but also an accurate and fully open and reusable source of bibliographic metadata to which members of the scholarly community can directly contribute.

\section{Acknowledgments}

We would like to thank Simone Persiani, Arianna Moretti, and Elia Rizzetto for their contributions to developing the OpenCitations Meta software.

\section{Funding information}

This work has been partially funded by the European Union's Horizon 2020 and Horizon Europe Research and Innovation Program under grant agreements No 101017452 (OpenAIRE-Nexus Project) and No 101095129 (GraspOS Project).

\section{Author contributions}
Arcangelo Massari: Data curation, Formal analysis, Investigation, Methodology, Software, Validation, Visualization, Writing – original draft, Writing – review \& editing. Fabio Mariani: Methodology, Software, Writing – original draft. Ivan Heibi: Writing – review \& editing. Silvio Peroni: Conceptualization, Funding acquisition, Methodology, Project administration, Supervision. David Shotton: Conceptualization, Project administration, Supervision, Writing – review \& editing.

\section{Competing interests}
The authors have no competing interests.

\printbibliography
\end{document}